\documentclass[twocolumn,showpacs,preprintnumbers,amsmath,amssymb]{revtex4}

\usepackage{graphicx}
\usepackage{dcolumn}
\usepackage{bm}

\begin{document}

\preprint{QDKD PRA resubmitted.tex}

\title{{\bf  Quantum dense key distribution }}

\author{I. P. Degiovanni}
 \email{degio@ien.it}
\author{I. Ruo Berchera}
\author{S. Castelletto}
\author{M. L. Rastello }
\affiliation{Istituto Elettrotecnico Nazionale G. Ferraris \\
Strada delle Cacce 91-10135 Torino (Italy)}

\author{F. A. Bovino }%
\email{fabio.bovino@elsag.it}
\author{A. M. Colla}
\author{G. Castagnoli}

\affiliation{%
ELSAG SpA \\
Via Puccini 2-16154 Genova (Italy)
}%

\date{\today}

\begin{abstract}

This paper proposes a new protocol for quantum dense key
distribution. This protocol embeds the benefits of  a quantum
dense coding and a quantum key distribution and is able to
generate shared secret keys four times more efficiently than BB84
one. We hereinafter prove the security of this scheme against
individual eavesdropping attacks, and we present preliminary
experimental results, showing its feasibility.
\end{abstract}


\maketitle



\section{Introduction}

Quantum dense coding (QDC) and quantum key distribution (QKD) are
two direct applications of fundamental quantum mechanics, both
involving two parties, Alice and Bob, exchanging some classical
information.

By QKD Alice and Bob exchange secret random keys to implement a
secure encryption-decryption algorithm (one-time pad) without
meeting, and the security of the distributed keys is based on the
laws of quantum physics \cite{gisinrevmod}.

Theoretically proposed in \cite{qdcpt}, QDC basically doubles the
capacity of transmission of a classical channel by local
operations on one particle of the EPR pair shared by the two
parties. QDC has received only partial experimental verification
\cite{qdcep} using polarization entangled states of photons
because of the inefficiently implemented Bell's state analysis
\cite{shihteleporting} and a lack of security in the transmitted
information.

Recently, Bostroem and Felbinger \cite{bostrom} gave birth to the
original idea of a protocol encoding secure information by local
operations on an EPR pair, though rather proposing  a
\textit{deterministic} and \textit{secure} transmission than
implementing a truly \textit{dense} coding scheme.

In this paper we propose a protocol for the quantum dense key
distribution (QDKD) including the advantages of QKD and QDC in
generating shared secret keys and enhancing transmission capacity.
In Section II we present the protocol, in Section III  we prove
the security of QDKD against the \textit{individual} eavesdropping
attack, in Section IV the experiment proves the QDKD feasibility,
and in Section V we discuss the practical efficiency of the
present protocol versus the most efficient QKD ones.

\begin{figure}[tbp]
\par
\begin{center}
\includegraphics[angle=0, width=7.5 cm, height=5 cm]{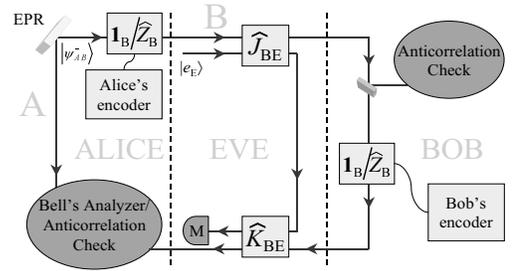}
\end{center}
\caption{ QDKD scheme: particle B of an EPR pair
$|\psi^{-}_{\mathrm{AB}}\rangle$ is sent from Alice to Bob and
backward. Alice's and Bob's encoders selecting the local operation
 on particle ($\mathbf{1}_{\mathrm{B}}$ or
$\widehat{Z}_{\mathrm{B}}$) together with the incomplete Bell's
measurement by Alice establish the key. Some pairs are randomly
selected for non-local measurement (the anticorrelation check
performed by spatially separated Alice and Bob measurements), as
test of the transmission security. The more general eavesdropping
attack is represented by the coupling of the EPR pair with a
larger Hilbert space in the initial state $|e_{\mathrm{E}}\rangle$
by means of unitary operators $\widehat{J}_{\mathrm{BE}}$ and
$\widehat{K}_{\mathrm{BE}}$.  } \label{Figure 1}
\end{figure}

\section{QDKD: the protocol}

The basic scheme of QDKD is presented in Fig. 1. Alice produces
pairs of particles in the singlet state
$|\psi^{-}_{\mathrm{AB}}\rangle=\frac{1}{\sqrt{2}}(|0_{\mathrm{A}}1_{\mathrm{B}}
\rangle-|1_{\mathrm{A}}0_{\mathrm{B}}\rangle)$, and stores
particle A in her lab, whereas she acts randomly with gate
$\mathbf{1}_{\mathrm{B}}$ or $\widehat{Z}_{\mathrm{B}}$ on
particle B and then sends it to Bob. As
$\widehat{Z}_{\mathrm{B}}|0_{\mathrm{B}}\rangle=|0_{\mathrm{B}}\rangle$,
and
$\widehat{Z}_{\mathrm{B}}|1_{\mathrm{B}}\rangle=-|1_{\mathrm{B}}\rangle$,
Alice's random selection of gate $\mathbf{1}_{\mathrm{B}}$ or
$\widehat{Z}_{\mathrm{B}}$ encodes the bits of her secret key on
the EPR pair,
\begin{eqnarray}
\mathbf{1}_{\mathrm{B}}|\psi^{-}_{\mathrm{AB}}\rangle&=&|\psi^{-}_{\mathrm{AB}}\rangle
\; \longrightarrow \; \mathrm{bit } \; 0  \nonumber \\
\widehat{Z}_{\mathrm{B}}|\psi^{-}_{\mathrm{AB}}\rangle&=&-|\psi^{+}_{\mathrm{AB}}\rangle
\; \longrightarrow \; \mathrm{bit } \; 1 , \label{a1}
\end{eqnarray}
with $|\psi^{+}_{\mathrm{AB}}\rangle=\frac{1}{\sqrt{2}}
(|0_{\mathrm{A}}1_{\mathrm{B}}\rangle+|1_{\mathrm{A}}0_{\mathrm{B}}\rangle)$.

Bob randomly switches particle B towards either his measurement or
his encoding apparatus. In one case Bob projects particle B on the
base $\{|0_{\mathrm{B}}\rangle, |1_{\mathrm{B}}\rangle\}$ while in
the other case, Bob, analogously to Alice, randomly acts with
$\mathbf{1}_{\mathrm{B}}$ or $\widehat{Z}_{\mathrm{B}}$ on
particle B and then sends it back to Alice. When she receives it,
her measurement apparatus performs an incomplete Bell's state
analysis, i.e. a projection on $|\psi^{+}_{\mathrm{AB}}\rangle$ or
$|\psi^{-}_{\mathrm{AB}}\rangle$ of the two-particle state
composed by the previously stored particle A and particle B. When
Bob's apparatus projects particle B, Alice's measurement apparatus
has to project particle A on the base $\{|0_{\mathrm{A}}\rangle,
|1_{\mathrm{A}}\rangle\}$ instead of performing a Bell's state
analysis. As Alice prepares only states
$|\psi^{-}_{\mathrm{AB}}\rangle$ and
$|\psi^{+}_{\mathrm{AB}}\rangle$, Alice and Bob results should be
always anticorrelated. The anti-correlation check in Fig. 1,
consists in comparing Alice and Bob results, and guarantees the
security of the distributed keys against \textit{individual}
eavesdropping attack \cite{gisinrevmod}.

After repeating this protocol enough times to produce the keys,
Alice discloses the results of her measurements
($|\psi^{-}_{\mathrm{AB}}\rangle$,
$|\psi^{+}_{\mathrm{AB}}\rangle$, $|0_{\mathrm{A}}\rangle$,
$|1_{\mathrm{A}}\rangle$) publicly exposing solely to the
possibility of being monitored but not modified. Alice's
measurements of $|\psi^{+}_{\mathrm{AB}}\rangle$ and
$|\psi^{-}_{\mathrm{AB}}\rangle$ indicates the generation of the
keys. In fact because Bob is aware of Alice's measurement results
(as well as of his own operation $\mathbf{1}_{\mathrm{B}}$ or
$\widehat{Z}_{\mathrm{B}}$), he can extract the bit encoded by
Alice and \textit{vice versa}. Specifically, measuring
$|\psi^{+}_{\mathrm{AB}}\rangle$ means that Alice and Bob encoded
0 and 1 or 1 and 0, respectively (Alice performed the operation
$\mathbf{1}_{\mathrm{B}}$ and Bob
 $\widehat{Z}_{\mathrm{B}}$ or \textit{vice
versa}). The measurement of $|\psi^{-}_{\mathrm{AB}}\rangle$,
instead, means that Alice and Bob encoded the same bit, 0 or 1
(Alice and Bob performed the same operations on particle B,
$\mathbf{1}_{\mathrm{B}}$ or $\widehat{Z}_{\mathrm{B}}$).

Thus, two keys are generated in this process, one produced by
Alice and one by Bob. The key generation is \textit{dense} because
any travelling particle B carries two exchanged bits, one
belonging to key A and one to key B.

Since Alice's results, disclosed publicly, correspond to the sum
(mod 2) of the bits of the two keys, i.e. $0\oplus0=1\oplus1=0$
and $0\oplus1=1\oplus0=1$, a possible eavesdropper (Eve) on the
public channel can intercept some information by correlating the
two keys, though she does not get information on the single key.
In other words, key B is used to encrypt key A and
\textit{viceversa}, and the results disclosed by Alice are the
"encrypted key A" sent to Bob. In fact both keys are used twice,
first to encrypt the other key, and then to encrypt Alice and Bob
messages.

This obviously induces a lack of security if the one-time-pad
protocol is implemented with both keys straightforwardly. For this
reason only key A is used directly in the implementation of the
one-time-pad protocol. The discussion on how Alice and Bob can
lower Eve's information on key B in order to generate a "new"
secret key B is a typical task in the field of computer science
and it is outside the scope of this paper.

Here we simply observe that Eve does not have any direct
information on the single key but only on the correlation between
the random bits in specific position of the two keys. If Alice and
Bob have common secret algorithms for the permutation of the
position of the bits of one of the two key (e.g. key B), they can
strongly reduce Eve's information on the two key. To guarantee the
complete security of the encrypted message we suggest to use only
key A for the one time pad, while the scrambled key B can be used
as a source for a cipher to generate a new key.

\section{QDKD: the security proof }

Let's consider now the \textit{individual} Eve's attack (Fig. 1)
consisting in coupling particle B with an \textit{ancilla} system
in the initial state $|e_{\mathrm{E}}\rangle$  by means of a
general unitary operator $\widehat{J}_{\mathrm{BE}}$. The Hilbert
space A-B is widened by coupling with the N-dimensional Hilbert
space E of the ancilla system. After Bob's operations, Eve couples
again her ancilla to system A-B by applying another unitary
operator $\widehat{K}_{\mathrm{BE}}$. Eventually she performs any
POVM measurement allowed by the laws of quantum mechanics on the
ancilla system E. Note that this approach is general because any
physical non-unitary interaction is equivalent to a unitary one
with a higher dimensional ancilla space \cite{gisinperes}. We
represent the coupled system A-B-E after Bob's and Eve's
operations as
\begin{eqnarray}
\widehat{K}_{\mathrm{BE}}\mathbf{1}_{\mathrm{B}}\widehat{J}_{\mathrm{BE}}
|\psi^{+}_{\mathrm{AB}}\rangle\otimes|e_{\mathrm{E}}\rangle&=&|\mu^{+}_{\mathrm{ABE}}\rangle,
  \nonumber \\
\widehat{K}_{\mathrm{BE}}\mathbf{1}_{\mathrm{B}}\widehat{J}_{\mathrm{BE}}
|\psi^{-}_{\mathrm{AB}}\rangle\otimes|e_{\mathrm{E}}\rangle&=&|\mu^{-}_{\mathrm{ABE}}\rangle,
  \nonumber \\
\widehat{K}_{\mathrm{BE}}\widehat{Z}_{\mathrm{B}}\widehat{J}_{\mathrm{BE}}
|\psi^{+}_{\mathrm{AB}}\rangle\otimes|e_{\mathrm{E}}\rangle&=&|\nu^{+}_{\mathrm{ABE}}\rangle,
  \nonumber \\
\widehat{K}_{\mathrm{BE}}\widehat{Z}_{\mathrm{B}}\widehat{J}_{\mathrm{BE}}
|\psi^{-}_{\mathrm{AB}}\rangle\otimes|e_{\mathrm{E}}\rangle&=&|\nu^{-}_{\mathrm{ABE}}\rangle.
\label{jkact}
\end{eqnarray}
As Alice prepares only states $|\psi^{\pm}_{\mathrm{AB}}\rangle$,
with probability $\frac{1}{2}$, the results of the
anti-correlation check performed by Alice and Bob should always be
anticorrelated. Eve's action induces a bit flip on particle B
giving correlated results. This is basically the signature of
Eve's presence.

We estimate now the probability $\mathrm{P}_{corr}$ of correlated
results in the anti-correlation check
\begin{equation}
  \mathrm{P}_{corr}=\mathrm{tr}[\widehat{J}_{\mathrm{BE}}\widehat{\rho}_{\mathrm{AB}}\otimes
  |e_{\mathrm{E}}\rangle\langle e_{\mathrm{E}}|\widehat{J}^{\dag}_{\mathrm{BE}}
  (\widehat{P}_{00}+\widehat{P}_{11})],     \label{pcorr1}
\end{equation}
where $\widehat{P}_{00}=|0_{\mathrm{A}}
0_{\mathrm{B}}\rangle\langle 0_{\mathrm{A}} 0_{\mathrm{B}}|$,
$\widehat{P}_{11}=|1_{\mathrm{A}} 1_{\mathrm{B}}\rangle\langle
1_{\mathrm{A}} 1_{\mathrm{B}}|$ are projection operators and
$\widehat{\rho}_{\mathrm{AB}}=1/2 |\psi^{-}_{\mathrm{AB}}\rangle
\langle \psi^{-}_{\mathrm{AB}} | + 1/2
|\psi^{+}_{\mathrm{AB}}\rangle \langle \psi^{+}_{\mathrm{AB}} |$.
From Eq.s (\ref{jkact}) and the properties of unitary operators
$\widehat{K}_{\mathrm{BE}}$ and $\widehat{J}_{\mathrm{BE}}$, it
can be demonstrated that
\begin{equation}
\mathrm{P}_{corr}= \frac{1}{2} (1+ \langle
\mu^{+}_{\mathrm{ABE}}|\nu^{-}_{\mathrm{ABE}}\rangle).
\label{pcorr2}
\end{equation}

The mutual information $I_{s:r}$ between the \textit{sender}
($s$), preparing and sending states $\widehat{\rho}_{n}$ with
probability $p_{n}$, and the \textit{receiver} ($r$) satisfies the
Holevo bound \cite{NC00}
\begin{equation}
I_{s:r}\leq \mathcal{I}_{\mathrm{s:r}}= S(\widehat{\rho})-\sum_{n}
p_{n} S(\widehat{\rho}_{n}), \label{hb}
\end{equation}
where $S(\widehat{\rho})$ is the Von Neumann entropy \cite{NC00}
of the generic state $\widehat{\rho}$, and
$\widehat{\rho}=\sum_{n} p_{n} \widehat{\rho}_{n}$. Note that,
even if the \textit{receiver} performs the more general POVM
measurements on the quantum state $\widehat{\rho}$, there is no
guarantee that he can reach the Holevo bound
$\mathcal{I}_{\mathrm{s:r}}$.

\begin{figure}[tbp]
\par
\begin{center}
\includegraphics[angle=0, width=7.5 cm, height=5 cm]{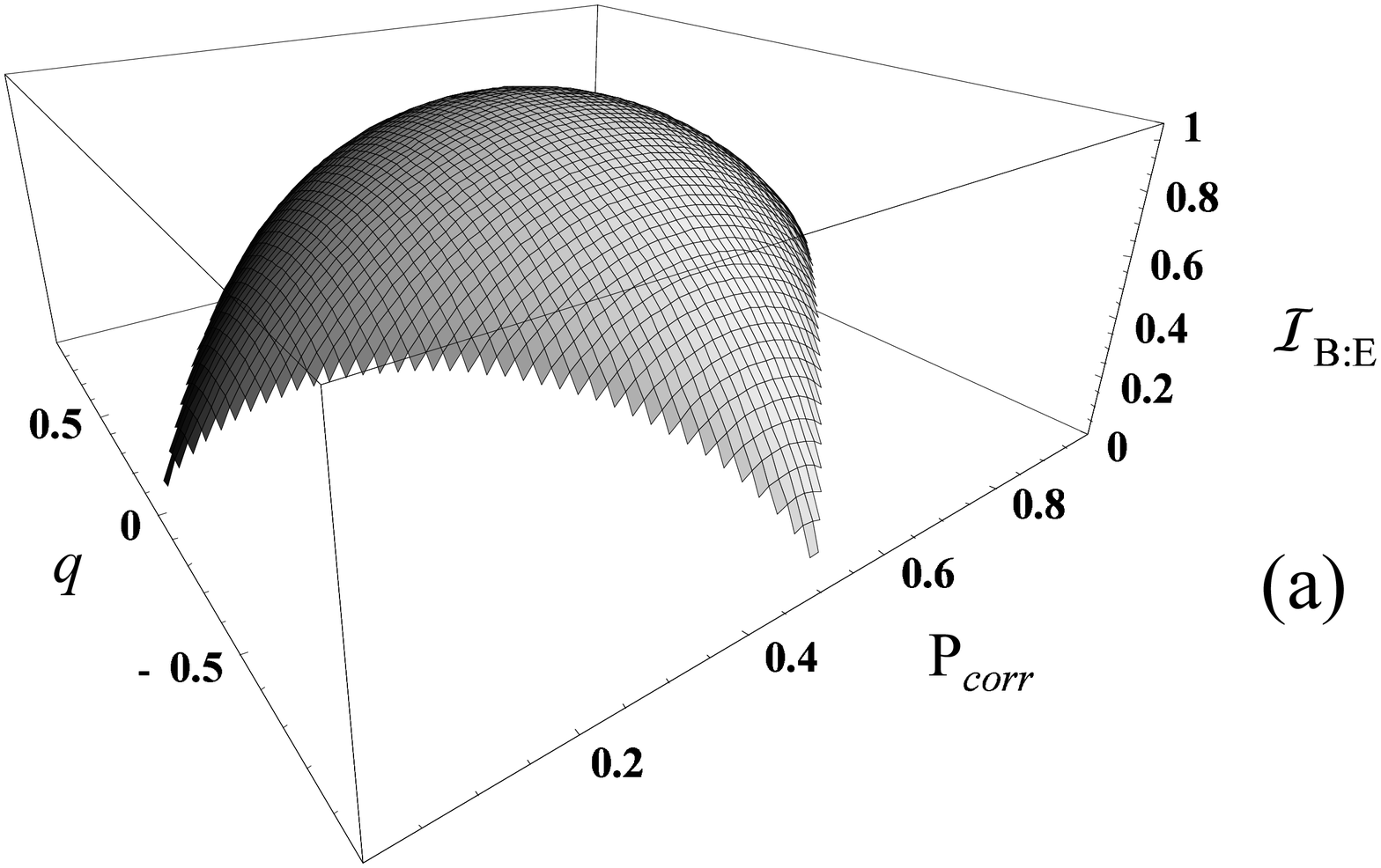}
\includegraphics[angle=0, width=7.5 cm, height=5 cm]{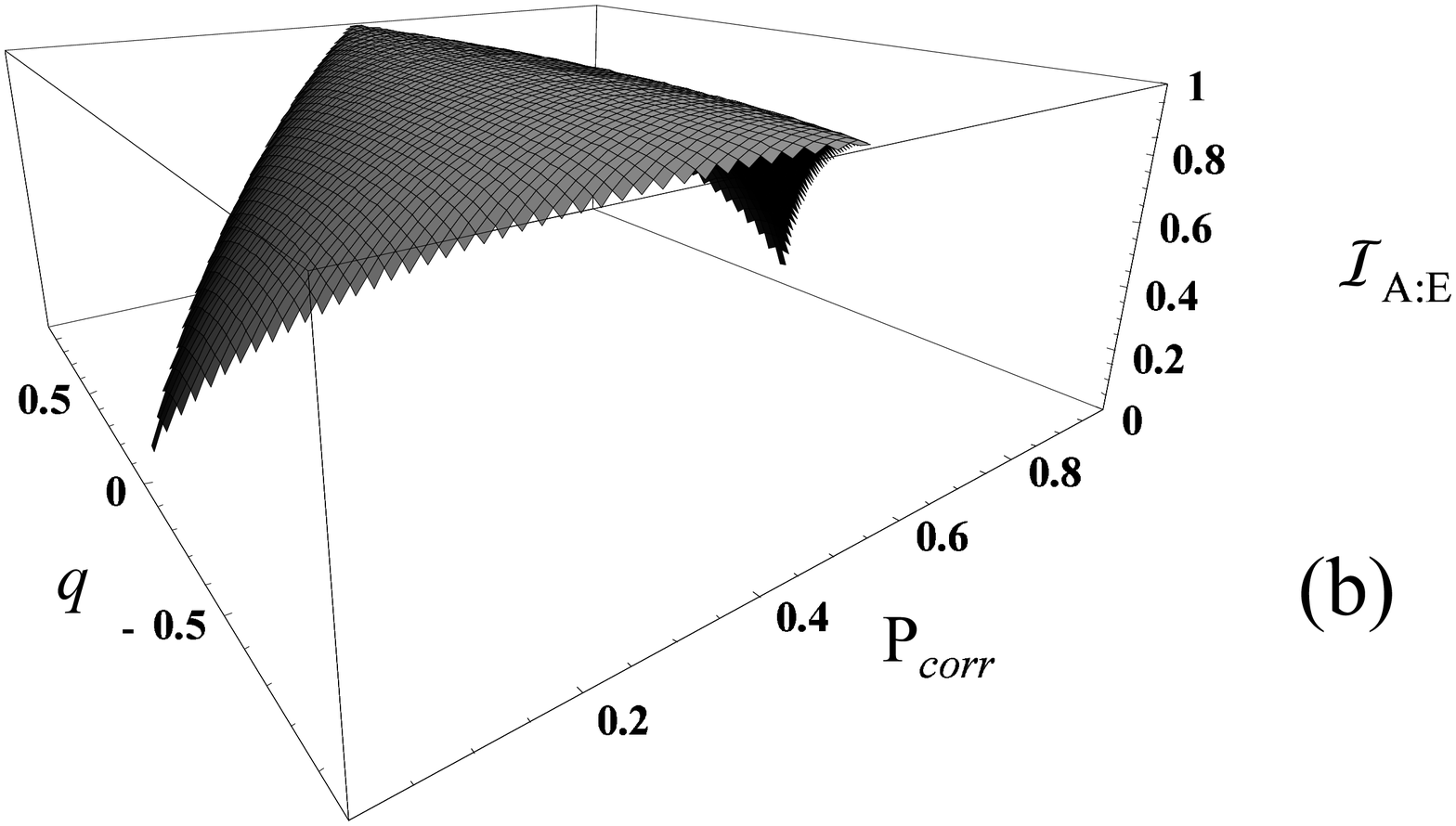}
\end{center}
\caption{ Plots of $\mathcal{I}_{\mathrm{B:E}}$ (a) and
$\mathcal{I}_{\mathrm{A:E}}$ (b) versus $\mathrm{P}_{corr}$ and
$q$. } \label{Figure 2}
\end{figure}

Following the same argumentation leading to Eq. (\ref{pcorr2}),
the Holevo bounds $\mathcal{I}_{\mathrm{B:E}}$ and
$\mathcal{I}_{\mathrm{A:E}}$ are expressed in terms of
$\mathrm{P}_{corr}$ and
$q=\langle\mu^{+}_{\mathrm{ABE}}|\nu^{+}_{\mathrm{ABE}}\rangle$.
Fig. 2 plots $\mathcal{I}_{\mathrm{B:E}}$ and
$\mathcal{I}_{\mathrm{A:E}}$ versus $\mathrm{P}_{corr}$ and $q$,
showing that the anti-correlation check provides an estimate of
the maximum information obtainable by Eve. From this estimate
Alice and Bob establish the security level of the bits exchanged.
This also implies that Eve cannot avoid to disclose herself when
attacking individually the qubit on the quantum channel.

Moreover Eve's possibilities have been overestimated in deriving
$\mathcal{I}_{\mathrm{B:E}}$ and $\mathcal{I}_{\mathrm{A:E}}$. In
fact Eve is assumed to perform any POVM on the final state of the
whole system A-B-E. This is obviously not the case, as she can
perform any POVM on the final state of her system E, but on system
A-B she only knows the results disclosed during the public
discussion.

In any practical implementation of QKD protocols, Alice and Bob
distill from the noisy and possibly insecure key a nearly
noise-free and secure key by means of error correction and privacy
amplification \cite{faint1}. Alice's and Bob's capability of
recovering two secure cryptographic keys (key A and key B) on the
face of Eve's presence, is given, though in some cases too
restrictively \cite{wolf}, by the condition $I_{\mathrm{A:B}}>
I_{\mathrm{A:E}}$ and $I_{\mathrm{A:B}}> I_{\mathrm{B:E}}$
\cite{ekertperes}, where $I_{\mathrm{A:B}}$ is the mutual
information between Alice and Bob.  $I_{\mathrm{A:B}}$ can be
simply calculated considering the capacity of a noisy channel of
quantum bit error rate $\mathcal{Q}$, as
$I_{\mathrm{A:B}}=1-H(\mathcal{Q})$, where $H$ is the Shannon
entropy of a binary channel \cite{shannon}. To ensure the security
of the two generated keys, we replace $I_{\mathrm{A:E}}$ and
$I_{\mathrm{B:E}}$ with the maximums of
$\mathcal{I}_{\mathrm{A:E}}$ and $\mathcal{I}_{\mathrm{B:E}}$
found in $q=0$, for any $\mathrm{P}_{corr}$. Regardless
differences in functions $\mathcal{I}_{\mathrm{A:E}}$ and
$\mathcal{I}_{\mathrm{B:E}}$, it has been demonstrated that
$\mathcal{I}_{\mathrm{A:E}}|_{q=0}=\mathcal{I}_{\mathrm{B:E}}|_{q=0}=H(\mathrm{P}_{corr})$
for any values of $\mathrm{P}_{corr}$. This means that Alice and
Bob can distill common secret keys when
\begin{equation}\label{dis2}
  H(\mathcal{Q})+H(\mathrm{P}_{corr})<1.
\end{equation}

\begin{figure}[tbp]
\par
\begin{center}
\includegraphics[angle=0, width=7.5 cm, height=5 cm]{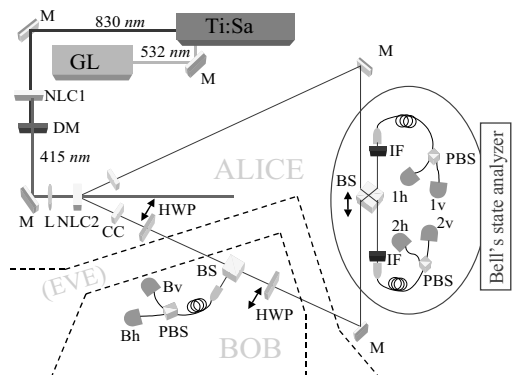}
\end{center}
\caption{ QDKD set-up: photon pairs are generated by SPDC in type
II nonlinear crystal (NLC2) pumped by the pulsed laser system (GL,
Ti:Sa and NLC1). The bits of the keys are encoded in the phase of
the entangled state by means of half-wave plates (HWPs). The
incomplete Bell's state measurement is performed with an
interferometer analogous to the one used in \cite{qdcep},
consisting of a 50:50 beam splitter (BS), interference filters
(IFs), fiber couplers, fibers integrated polarizing beam splitters
(PBS), single-photon detectors. M stands for mirror, DM dichroic
mirror, L lens, CC birefringence compensation crystals.  }
\label{Figure 3}
\end{figure}

\section{QDKD: the experiment}

We implemented experimentally a proof of principle of QDKD
protocol by the scheme of Fig. 3.

The source of entangled photons in a singlet state is a 0.5 mm
long BBO nonlinear crystal (NLC2) pumped by ultrashort pulses (150
fs) at 415 nm from a second harmonic (obtained from NLC1) of a
ultrashort mode-locked Ti-Sapphire with repetition rate 76 MHz
pumped by a 532 nm laser (GL). NLC2 realizes a non-collinear type
II phase matching, to obtain polarization entangled state
$|\psi^{-}\rangle$ at 830 nm, where $|0\rangle$ [$|1\rangle$]
stands for horizontal [vertical] polarization \cite{kwiat2}.
Alice's bit encoding procedure consists in taking on and off, in
the quantum channel B, an half-wave plate (HWP) with optical axis
parallel to the horizontal polarization state. According to Eq.
(\ref{a1}) the  HWP on means Alice's encoding bit 1, while
 off corresponds to bit 0. The same holds for Bob's
encoding apparatus. The Bell's state analyzer is a Hong-Ou-Mandel
interferometer \cite{Hong-Ou-Mandel} together with two
polarization-measurement systems. To compensate for the optical
delays in the interferometer the 50:50 beam splitter (BS) is
mounted on a micrometric translation stage. The anti-correlation
check consists in polarization measurements performed separately
by Alice and Bob on photons A and B respectively, when photon B is
reflected by Bob's BS towards his detection apparatus. As a matter
of fact Bob decides the ratio of pairs devoted to this measurement
since the reflectivity of his BS sends randomly some photons B
towards his polarization measurement system. Note that Alice's
Bell-state-analyzer is also suitable for polarization measurement,
namely for the anti-correlation check. Alice's and Bob's detection
apparatuses are composed by open air-fiber couplers to collect the
down-converted light in single-mode optical fibers. The detection
of photons according to their polarization is guaranteed by a
fiber-integrated polarizing beam splitter (PBS). Photons at the
two output ports of PBS are sent to fiber coupled photon counters
(Perkin-Elmer SPCM-AQR-14). These are indicated as 1h, 1v, 2h, and
2v in the Alice's detection apparatus, and as Bh and Bv in the
Bob's anticorrelation check. Interference filters (IF, at 830 nm,
11 nm FWHM) are placed in front of the fiber couplers to reduce
stray-light.

Fig. 4 shows the interference profiles when both Alice and Bob
encode bit 0 (a), Alice encodes bit 1 and Bob bit 0 (b), both
Alice and Bob encode bit 1 (c). Cases (a) and (c), corresponding
to a final state $|\psi^{-}_{\mathrm{AB}}\rangle$, present peaks
in the coincidence counts from detectors 1h 2v or 1v 2h and dips
in the coincidence from detectors 1h 1v or 2h 2v, since
$|\psi^{-}_{\mathrm{AB}}\rangle$ is spatially antisymmetric. By
contrary case (b), corresponding to the spatially symmetric final
state $|\psi^{+}_{\mathrm{AB}}\rangle$, presents the complementary
behavior. When the paths of photons A and B differ more than their
coherence length, no interference occurs, and one obtains
classical statistics for the coincidence count rates at the
detectors. For optimal positioning of BS, i.e. indistinguishable
photon paths, interferences enables Alice and Bob to read the
encoded information. We measured $\mathrm{P}_{corr}$ as the ratio
between the sum of coincidences from detectors Bh 1h and Bv 1v
corresponding to correlated results, and the sum of coincidences
between Bh 1h, Bv 1v, Bh 1v, and Bv 1h, corresponding to both
correlated and anticorrelated results. We obtained
$\mathrm{P}_{corr}<0.05$. The presence of correlated counts
together with the reduction of the visibility of the
interferometric profiles are basically due to losses in optics,
detection, electronics, and noise \cite{pranostro2}. From the
visibility in Fig. 4 we estimated $\mathcal{Q}=3.3\%$, and
according with Eq. (\ref{dis2}) $H(0.033)+H(0.05)=0.49<1$.

\section{Further discussion}

According to Ref. \cite{cabello}, the theoretical efficiency of a
QKD protocol is defined as $\mathcal{E}=b_{s}/(q_{t}+b_{t})$,
where, for every step, $b_{s}$ is the expected number of secret
bits exchanged by the two parties, $q_{t}$ is the number of qubits
exchanged on the quantum channel, and $b_{s}$ is the number of
bits exchanged on the public channel. Any QKD protocol satisfies
$\mathcal{E}\leq 1$, while DQKD reaches the limit value
$\mathcal{E}= 1$ as $b_{s}=2$, $q_{t}=1$ and $b_{t}=1$, and the
bits used for eavesdrop checking are neglected \cite{cabello}.

To our knowledge only two other protocols reach the limit value of
$\mathcal{E}= 1$, one proposed by Cabello (high capacity Cabello
protocol, HCCP) \cite{cabello}, and one by Long and Liu (high
capacity Long Liu protocol, HCLLP) \cite{long}. Both protocols
exploit the fact that a possible eavesdropper, with no access to
the whole quantum system at the same time, cannot recover the
whole information without being detected, and both employ a larger
alphabet, a four dimensional orthogonal basis of pure states.

HCLLP appears hardly implementable as the detection system should
perform a complete discrimination between the four Bell's states
\cite{shihteleporting}. As to HCCP the detection system is
completely analogous to the interferometer implemented in the DQKD
experiment. On the contrary, in DQKD the source of photons and the
encoding apparatuses are quite simple (a type II parametric
down-conversion (PDC) source and two local operations), while they
are much more complicated in HCCP, where they could be implemented
by using a system to switch between two PDC sources (one type I,
and one type II) and three local operations. In conclusion DQKD is
a new maximally efficient QKD protocol fully exploiting QDC
instead of a larger alphabet, and incorporating the practical
advantages of HCCP and HCLLP.

In order to evaluate the effective potential applications of DQKD
we analyze the effects of losses, considering the length $L$ of
the communication channel between Alice and Bob. As the photon
sent from Alice to Bob and backwards travels over a distance $2L$,
the other photon of the pair should be stored by Alice for a time
$2L/c$ ($c$ is the speed of light in the quantum channel), for
example in a storage fiber ring of length $2L$. Thus, for DQKD the
two photons should be kept decoherence-free for a time $2L/c$, as
for HCCP, while for HCLLP they should be kept decoherence-free for
a time $3L/c$.

In this respect we define a probability $\mathcal{P}$ that a
photon is transmitted over the distance $L$ \cite{nota}. The
practical efficiency, $\eta$, of a QKD protocol can be evaluated
by the product between the theoretical efficiency $\mathcal{E}$
and the probability that the photons are not lost in the
communication channels. In the case of DQKD (as well as of HCCP)
$\eta=\mathcal{P}^{2}\mathcal{P}^{2}$, because both photons travel
for a time $2L/c$. In the case of BB84 protocol
\cite{bennet&brassard} $\eta=0.25 \mathcal{P}$, as there is only
one photon travelling for the path $L$, and the theoretical
efficiency is only $\mathcal{E}=0.25$. An optimized version of the
BB84 exploits the possibility that Bob measures the photon sent by
Alice only after Alice has disclosed the measurement basis. In
this case $\mathcal{E}=0.5$, but after travelling for the distance
$L$, the photon should be stored by Bob for an additional time
$L/c$, waiting for Alice classical bits, resulting $\eta=0.5
\mathcal{P}^{2}$.

\begin{figure}[tbp]
\par
\begin{center}
\includegraphics[angle=0, width=7.5 cm, height=5 cm]{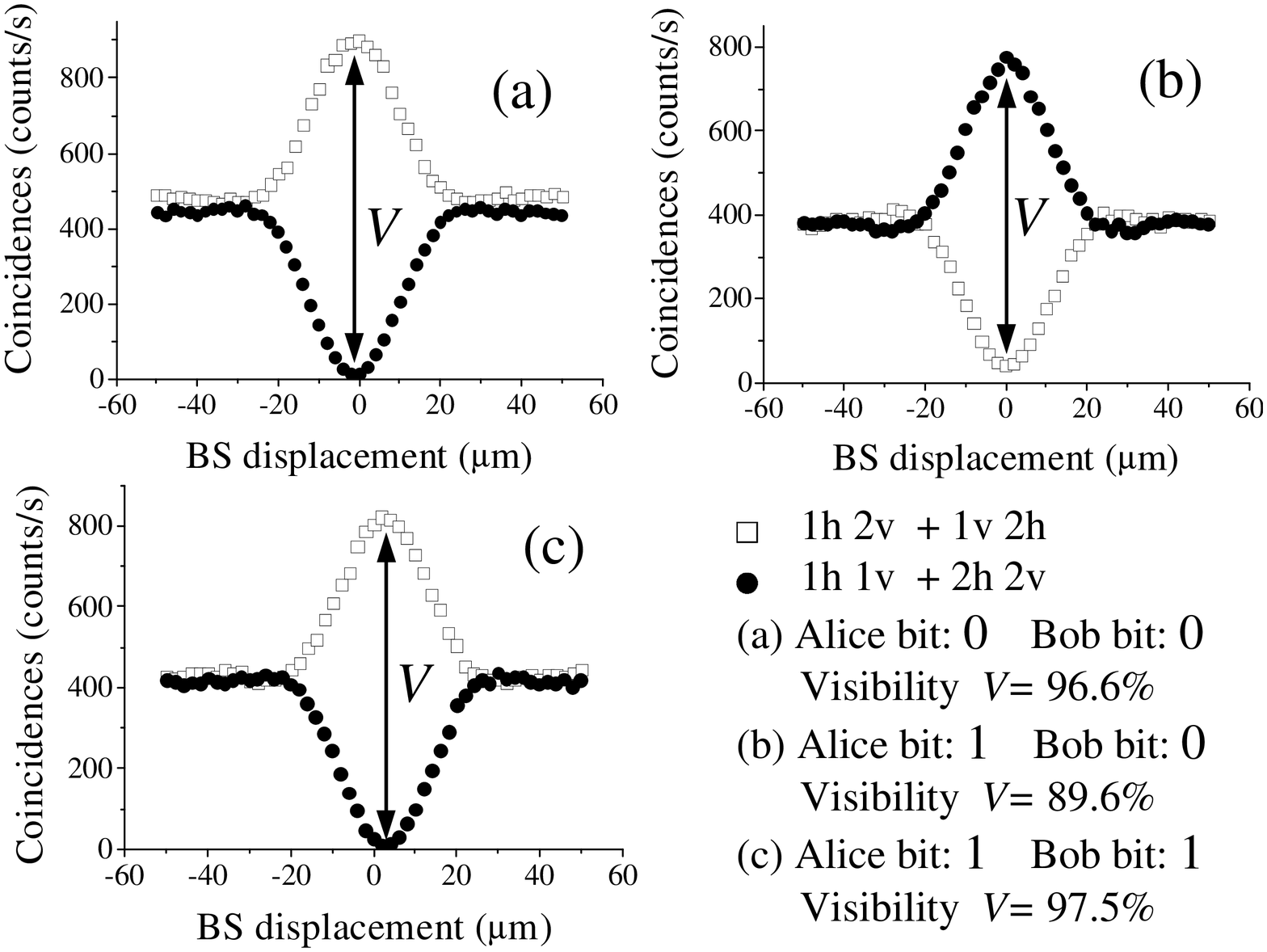}
\end{center}
\caption{Coincidence rates obtained from pairs of detectors 1h, 2v
or 1v, 2h (open square), and 1h, 1v or 2h, 2v (circle) versus the
displacement of the beam splitter. } \label{Figure 4}
\end{figure}

Eventually, the practical efficiency of a QKD protocol depends on
the quality of the quantum channel: if $\mathcal{P}\geq 71 \%$ the
most efficient protocols is still DQKD (as well as HCCP), if $50
\% \leq\mathcal{P} < 71 \%$ the more efficient is the optimized
BB84, while if $\mathcal{P} < 50 \%$ the best performance are
obtainable by the original BB84 protocol. For example, if we
consider 1550 nm commercial telecommunication fiber, whose
attenuation factor is 0.2 dB$/$km, DQKD protocol is more efficient
than BB84 protocols for distance $L$ shorter than 7.4 km, for
$L>15$ km the most efficient is the original BB84, while for
intermediate distances the more suitable choice is the optimized
BB84. However, the forthcoming advent of a new generation of
optical fibers, such as photonic-cristal fibers (PCF) and Bragg
fibers with predicted attenuation factors between $10^{-2}$
dB$/$km and $10^{-4}$ dB$/$km or less, will lead  DKQD protocol
advantageous over much longer distances, namely up to $L\simeq
150$ km corresponding to attenuation of $10^{-2}$ dB$/$km and up
to $L\simeq 15000$ km for attenuation of $10^{-4}$ dB$/$km.

\section{Conclusion}

In conclusion the proposed protocol embeds the main advantages of
two quantum communication applications, namely QKD and QDC, as it
allows the generation of secure cryptographic keys using only one
travelling-qubit for two bit of classical information.

The protocol has been proved to be secure against individual
eavesdropping attack by using a non-local measurement. We proved
also that QDKD reaches the maximum of the efficiency of QKD
protocols.

We performed a  experimental implementation establishing the
conditions for a secure QDKD. The experiment in fact proved the
feasibility of QDKD by showing high-visibility interferometric
profiles as well as low-noise anticorrelation check. This is only
the proof of principle of the system described above for QDKD, as
we are aware that the actual system is not yet fast enough in
switching the encoding operations. We are now focused on the
realization of fast switching $\widehat{Z}_{\mathrm{B}}$ gates by
using Pockell's cells and HWPs.

\begin{acknowledgments}

We are indebted with M. Genovese, P. Varisco, A. Martinoli, P. De
Nicolo, and S. Bruzzo. The work was carried out within a project
entitled "Quantum Cryptographic Key Distribution" co-funded by the
Italian Ministry of Education, University and Research (MIUR) -
grant n. 67679/ L. 488.
\end{acknowledgments}

\newpage

\vspace{5.5 cm}

\newpage

\vspace{5.5 cm}

\newpage

\vspace{5.5 cm}

\newpage

\vspace{5.5 cm}


\begin{thebibliography}{99}
\bibitem{nota}  We assume a typical exponential attenuation law, thus $\mathcal{P}=\exp(-L/\mathcal{L})$ where $\mathcal{L}$
is the characteristic attenuation length of the quantum channel.
\end{thebibliography}
\end{document}